# *Soft and chiral phonons in chiral phase of $K_3NiO_2$*

*Hong Dang Nguyen[1,2], Fedir Borodavka[1], Miroslav Lebeda[1,3], Nazar Zaremba[4], Peter Höhn[4], , Eteri Svanidze[4], Jan Drahokoupil[1,2], Aleš Vlk[1], Stanislav Kamba[1,*]*

*[1]Institute of Physics, Czech Academy of Sciences, Na Slovance 2, 182 00 Prague 8, Czech Republic*

*[2]Faculty of Nuclear Sciences and Physical Engineering, Czech Technical University in Prague, Trojanova 13, 120 00, Prague 2, Czech Republic*

*[3]Faculty of Mechanical Engineering, Czech Technical University in Prague, Technická 4, 16607 Prague 6, Czech Republic*

*[4]Max-Planck-Institute for Chemical Physics of Solids, Nöthnitzer Str. 40, 01187 Dresden, Germany*

**Corresponding author:**

Stanislav Kamba
kamba@fzu.cz

**Abstract**

Raman scattering measurements confirmed the theoretical prediction that the structural phase transition from the achiral tetragonal to the chiral tetragonal phase, which occurs near 400 K, is induced by a doubly degenerate soft phonon at the *Z* point of the Brillouin zone. In the low-temperature chiral phase, the soft mode activates in Raman spectra, splits into two components with $A_1$ and $B_1$ symmetries and harden with cooling according to Cochran's law. Circularly polarized Raman scattering did not reveal the angular momentum of these singly degenerate phonons at the Γ point, which is consistent with theory. We also calculated the phonon branches in the whole Brillouin zone for both crystalline phases and compared the results with the phonons observable in the Raman spectra. The calculations revealed that some phonons with $\boldsymbol{k} \neq 0$ have angular momentum in the chiral phase. A pronounced circular motion of atoms can be observed, for example, in a Dirac-type topological phonon at the M-point of the Brillouin zone with a frequency of 168 cm$^{-1}$.

## 1. Introduction

Recently, research on chiral materials and their excitations, such as phonons and magnons, has become a focal point in the field of solid-state physics[1]. A chiral structure is characterized by the absence of improper rotation; that is, it is impossible to overlap ~~a~~ right- and ~~a~~ left-handed structures through rotations and translations. A chiral crystal structure may promote the formation of chiral phonons with angular or magnetic momentum.

$K_3NiO_2$ is a very promising chiral system warranting further research. At high temperatures, $K_3NiO_2$ crystallizes in an achiral tetragonal structure with the space group $P4_2/mnm$ (No. 136, $Z$=2)[2,3]. Upon cooling below 423 K, $K_3NiO_2$ undergoes a structural phase transition to a chiral phase with twice larger unit cell ($Z$ = 4). This results in a racemic twin, consisting of the $P4_12_12$ and $P4_32_12$ chiral tetragonal phases, with both enantiomorphs having approximately the same volume. Between 400 and 423 K, the coexistence of the high-temperature and low-temperature phases was observed, indicating the phase transition is first-order [2,3]. Previous magnetic studies revealed that $K_3NiO_2$ is paramagnetic throughout the entire studied temperature range, down to 2 K [2], which was also confirmed as part of this work – see Supplementary Information (SI).

It has recently been theoretically suggested that the handedness of the $K_3NiO_2$ structure can be controlled by an external circularly polarized electromagnetic field [4]. Furthermore, Fava *et al.*[5] investigated the structural phase transition in $K_3NiO_2$ using first-principles calculations and they concluded that the low-temperature chiral phase is likely induced by an unstable, doubly degenerate soft phonon, which is located at the $Z$-point [$\boldsymbol{k}$ = (0, 0, 1/2)] of the Brillouin zone (BZ) in the high-temperature achiral phase. Due to the folding of the BZ in the chiral phase, this phonon moves to the Γ point and should become Raman-active and harden with cooling.

The goal of our work is to confirm this theoretical prediction and to test the possible chirality of the observed phonons using circularly polarized Raman scattering. In our Raman spectra, we ultimately observed two achiral soft modes. We then compared the experimental results with the results of our calculations of crystal lattice dynamics, which showed that the soft modes are singly degenerate and therefore cannot be chiral. However, the calculations confirmed the circular polarization of a series of phonons with nonzero wave-vectors.

## 2. Experiments

In contrast to the azide-nitrate method preferred in the literature [3], we developed a new synthetic route to $K_3NiO_2$ that proceeds without azide decomposition and thus without pressure, making it not only significantly safer, but also yielding better preparative results in terms of sample homogeneity and crystal quality and finally also faster.

In a first step, $K_2O$ was prepared by mixing potassium metal (Alfa Aesar, 99,95 %) and $KNO_3$ (Sigma Aldrich, 99,999 %) in a ratio of 5.5:1. The mixture was heated in a nickel crucible (Bochem) within a glove box under Ar to 300 °C; a light gray powder was obtained after a rapid and very violent reaction. The product was subsequently heated to 200 °C under vacuum for 2 hours to remove any excess potassium metal. The resulting ultrafine homogeneous powder showed the peaks of $\gamma$-$K_2O$ [6] only in the X-ray powder diffraction pattern and is highly sensitive to moisture.

To prepare $K_3NiO_2$, in a second step, K, $K_2O$ and NiO (Alfa Aesar, 99,998 %) in molar ratios between 2:1:1 and 6:1:1 were mixed, pelletized and placed in a massive Ni Ampulla sealed with a press-fit. For heat treatment, the sample was heated to 400 °C at a rate of 100 °C/h in a static argon atmosphere, held at that temperature for 48 hours, and then cooled to room temperature at a rate of 10 °C/h. The product consisted of a gray, ductile mass with carmine-red inclusions.

Without further preparation, the sample was heated in a third step at a rate of 300 °C per hour under dynamic vacuum ($\sim 1.10^{-5}$ mbar) to 300 °C, held at this temperature for 8 hours to remove potassium metal, and then cooled to room temperature at a rate of 300 °C/h. The resulting product was microcrystalline, reddish-brown in color, and consisted almost exclusively of $K_3NiO_2$ with very small amounts ($\ll$ 10 %) of $K_2NiO_2$ [7]. Although repeated recrystallization attempts with additional potassium metal improved the purity of the sample, the intermediate homogenization steps necessary resulted in significantly smaller crystallites that were largely unusable for our investigations. Increasing amounts of potassium metal in the starting mixture seemed to slightly improve crystal quality and size; to prepare the sample shown in Figure 1, a mixture of 129,5 mg K, 45,4 mg $K_2O$ and 36,4 mg NiO was employed.

The crystalline grains were reddish brown in color and had a diameter of about 10-20 micrometers. They degraded rapidly when exposed to air, so we stored and measured them directly in Argon-filled quartz capillaries with a diameter of 0.5 mm or 1.0 mm and a length of 4 cm. A photograph of the powdered samples is shown in Fig. 1.

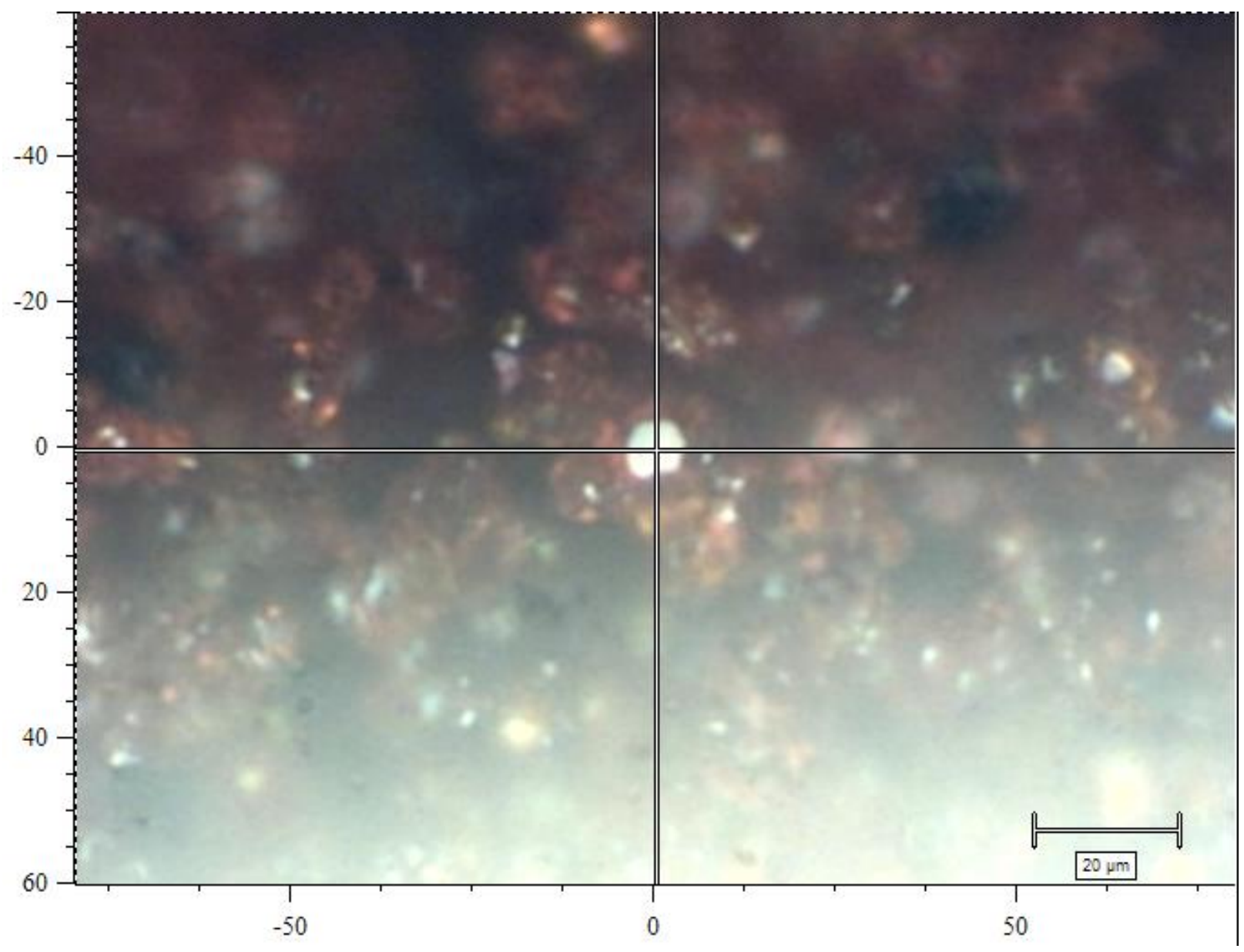

FIG. 1. Photo of $K_3NiO_2$ grains obtained using Raman optical microscope. White spot shows focus of Raman laser.

For Raman studies, a Renishaw RM 1000 Micro-Raman spectrometer equipped with a CCD detector and Bragg filters was used. The experiments were performed in the backscattering geometry within the 5–1000 $cm^{-1}$ range using a He-Ne laser with the wavelength of 633 nm and a Linkam THMS 600 temperature cell. The spectra were carefully fitted with a sum of independent damped harmonic oscillators. For circular polarization measurements a set of quartz zero-order achromatic quarter-wave plates was used.

**Calculations** with universal machine learning interatomic potential (uMLIP) were performed within the Atomic Simulation Environment (ASE)[8], using the foundation model of PET-MAD-1.5 (S) [9,10]. The calculation scripts were generated with the uMLIP-Interactive interface [11]. Before the phonon calculations, atomic positions and lattice vectors were relaxed with the LBFGS algorithm while keeping the lattice angles fixed until the maximum force was below 0.005 eV/Å. Phonon properties were calculated with Phonopy[12,13], using the finite-displacement method with a displacement amplitude of 0.01 Å. Phonon dispersions in the whole BZ were evaluated along the high-symmetry path Γ-X-M-Γ-Z-R-A-Z | X-R | M-A. A supercell of 3 × 3 × 2 (432 atoms) was used for the lower-temperature $K_3NiO_2$ phase (#92, 24 atoms), whereas 3 × 3 × 3 (324 atoms) was utilized for the higher-temperature $K_3NiO_2$ phase (#136, 12 atoms). For comparison, we also performed some calculations using the SevenNet-Nano model.

## 3. Results and discussion

### 3.1 Temperature-dependent and linearly polarized Raman spectra

Raman spectra were measured for both parallel (VV) and perpendicular (HV) polarization of the incident and scattered laser beams. The HV-polarized spectra had much lower intensities than the VV-polarized spectra, but their shapes were nearly identical due to random orientation of the crystals (see Fig. 2). Therefore, we present here only the temperature dependence of the VV-polarized spectra. Although the spectra were measured up to 1000 cm$^{-1}$, we observed most of phonons below 200 cm$^{-1}$. Only one additional strong mode was observed near 680 cm$^{-1}$. Since we focused on finding a soft phonon below 200 cm$^{-1}$, we did not study the detailed temperature dependence of this high-frequency mode. The low-temperature and high-temperature spectra obtained on different grains, are shown in Figs. 2 and 3. We fitted the spectra using a sum of harmonic oscillators,

$$I(\omega, T) = [1 + n(\omega, T)] \sum_j \frac{A_j \Omega_j^2 \Gamma_j \omega}{\left(\Omega_j^2 - \omega^2\right)^2 + \Gamma_j^2 \omega^2} \quad (1)$$

where $n(\omega, T)$ is the Bose-Einstein factor and $A_j, \Omega_j, \Gamma_j$ are the strength, wavenumber, and damping coefficient of the $j$-th oscillator, respectively. The obtained temperature dependence of the phonon frequencies is shown in Fig. 4. The following characteristics are observed:

1. Most of phonons exhibit a decrease in frequency upon heating. This can be explained by the anharmonic behavior of the crystal lattice vibrations[14,15]. Only two modes visible near 140–150 cm$^{-1}$, show an apparent increase in frequency at high temperatures. This is due to the high attenuation of these modes and, consequently, the high uncertainty in determining their frequencies. In fact, the damping of all phonons also increases with temperature due to anharmonic processes.[14]

2. At frequencies below 40 cm⁻¹, a doublet is observed at 80 K. Both modes soften significantly with heating in a manner consistent with Cochran's law[16,17]

$$\omega_{\mathrm{SM}}(T) = [A\,(T_c - T)]^{1/2}, \qquad (2)$$

for soft modes with a critical temperature of $T_c$ = 420 K (see Fig. 5). The soft phonons disappear from the spectra already near 400 K. This is likely caused by a first-order phase transition, during which the soft mode in the Raman spectrum is suddenly deactivated and the material rapidly transforms into a high-temperature achiral phase, as has been observed previously in X-ray diffraction [2,3]. It is also possible that the actual temperature of the sample was higher than what the thermometer in the Linkam cell indicated, due to the crystals being heated by the laser.

3. Both soft modes overlap at temperatures above 200 K due to the increase in their damping and their mutual overlap.

4. We measured dozens of crystal grains at room temperature, and the soft modes were visible only in some of them. This is because the soft modes are active only when measuring light scattering from the tetragonal plane. If the grains were oriented differently, the soft modes were not active.

5. In some cases, we detected only one soft mode across the entire temperature range studied (see sample #1 in Fig. 5). An explanation for the existence of one or two soft phonons will be presented below. A single soft mode theoretically exhibits a higher phase transition temperature ($T_c$ = 507 K). There are two possible explanations for this: (a) The phonon frequency is determined with greater uncertainty due to higher phonon damping, resulting from the overlap of two soft phonon components. For that reason, extrapolated $T_c$ is obtained also with higher uncertainty. (b) The theoretical $T_c$ is approximately 80 K above the actual phase transition temperature due to the first-order nature of the phase transition near 420 K [18].

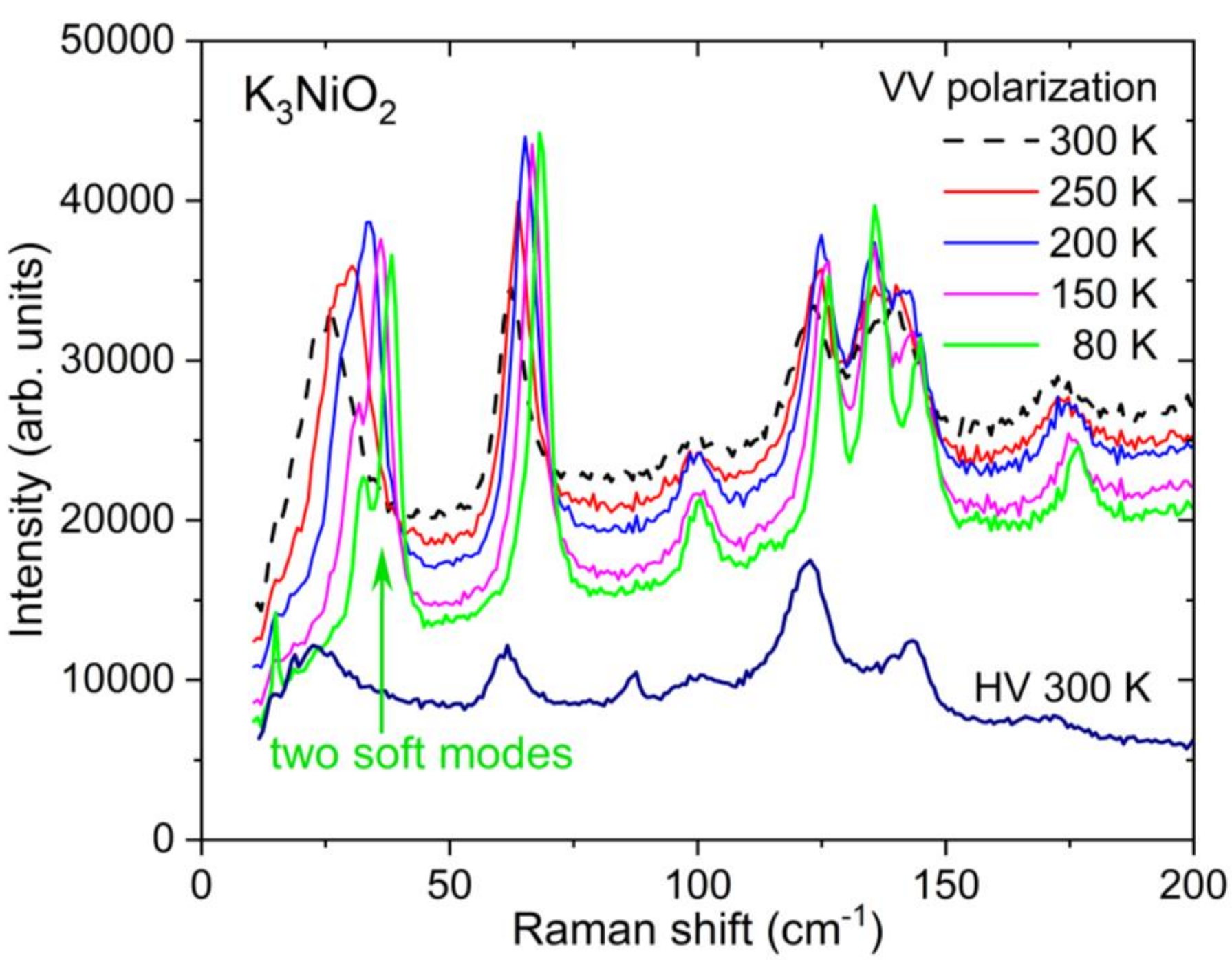

FIG. 2. Temperature dependence of parallel-polarized (VV) Raman spectra. Note the soft mode doublet below 40 $cm^{-1}$ seen at lowest temperatures. For comparison, the room temperature crossed-polarized (HV) spectrum is also shown. It has the same shape as VV spectrum but a much lower intensity.

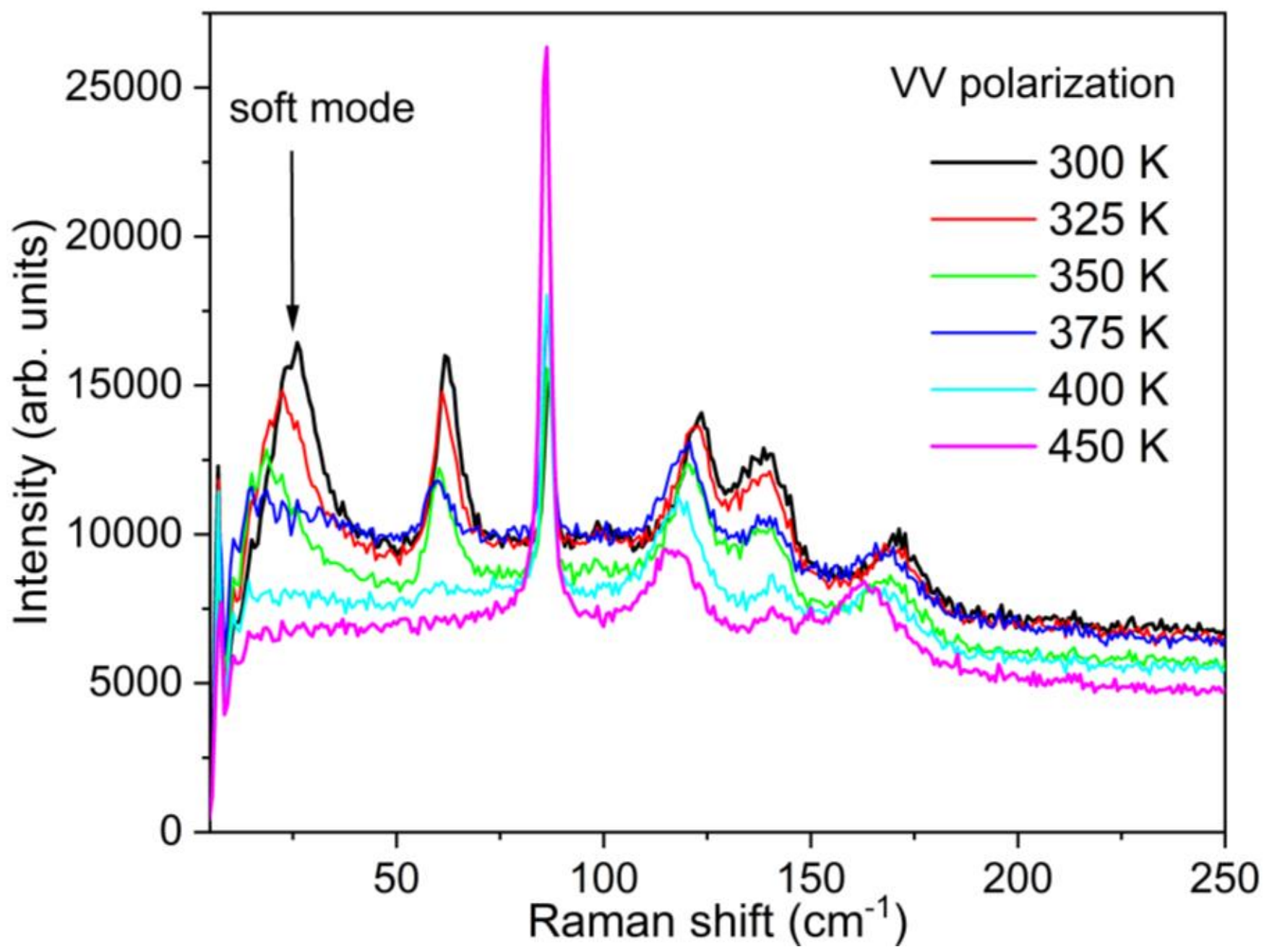

FIG. 3. High temperature Raman spectra of $K_3NiO_2$ measured in parallel polarization. The soft phonon disappears already at 400 K. The sharp peak at 90 $cm^{-1}$ is not visible in Fig. 2 because both measurements were performed on different crystals with different orientations.

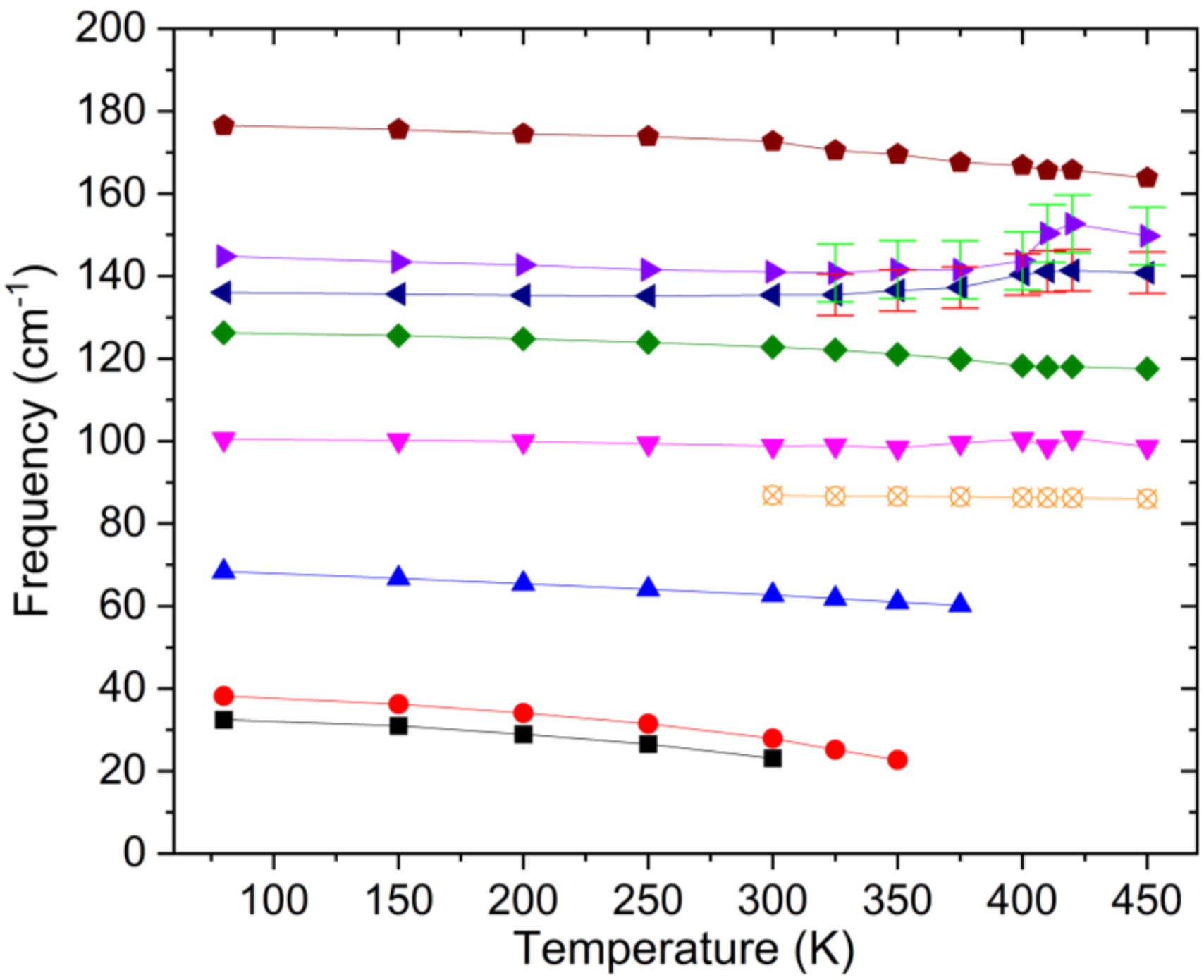

FIG. 4. Temperature dependence of phonon frequencies obtained from the fits of Raman spectra using Eq. (1). For most modes, the uncertainty of determination of their frequencies is smaller than the symbol size. Only two overlapped modes near 140–150 cm$^{-1}$ have due to high damping a higher uncertainty above room temperature, as indicated by the error bars.

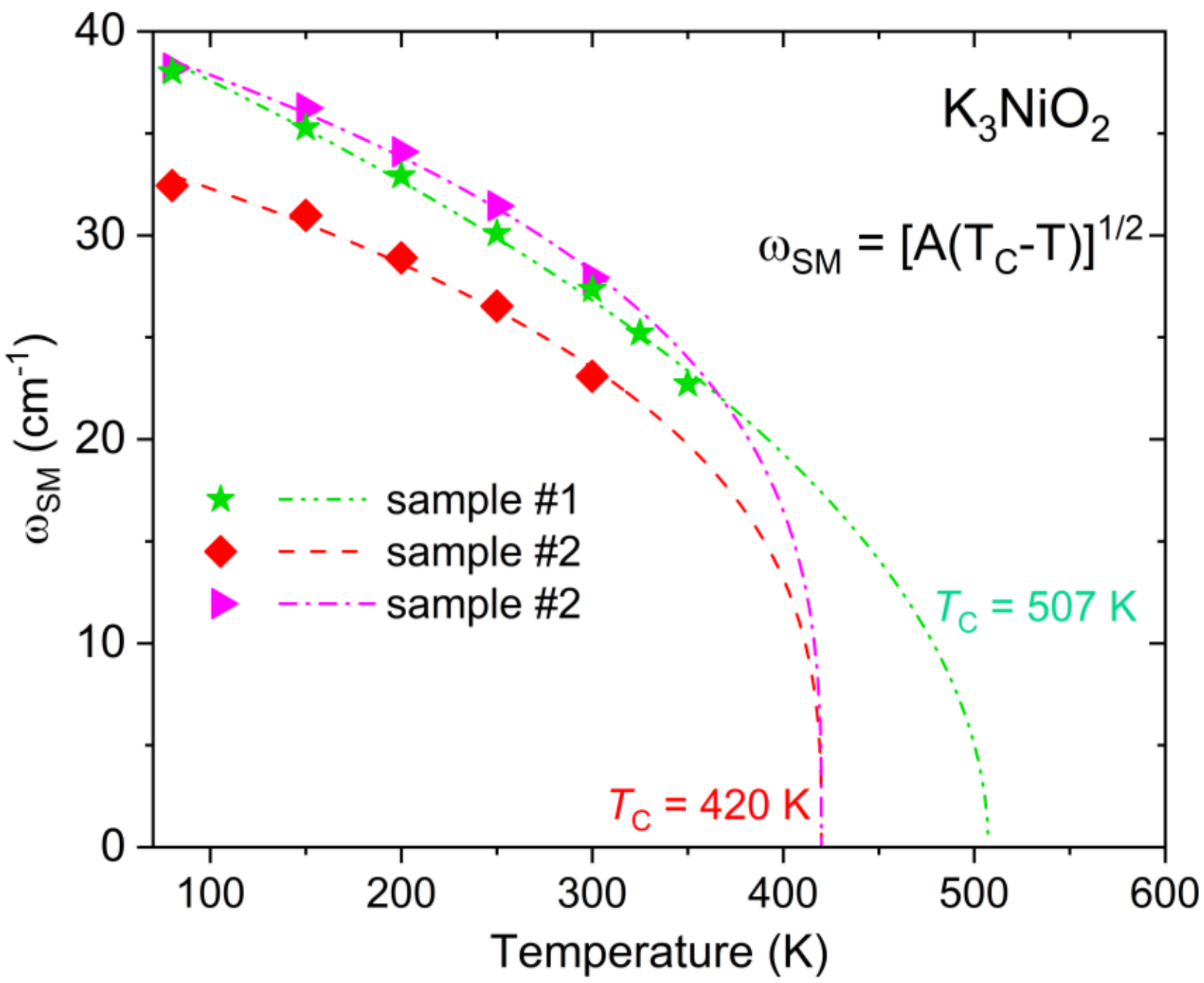

FIG. 5. Temperature dependence of the soft phonon frequencies. Sample #1 exhibits only single soft phonon due to improper orientation of the crystal, while sample #2 revealed soft phonon doublet. Details are explained in the text.

To better understand Raman spectra, we performed a factor-group analysis of phonon symmetries at the Γ point of BZ and evaluated their theoretical activity in Raman and infrared spectra. In the high-temperature *achiral phase above 420 K*, where the material crystallizes in a tetragonal $P4_2/mnm$ structure with two formula units per unit cell, the following *optical* phonons are allowed:

$$\Gamma_{opt} = 3A_{2u}(z) + 3B_{1u}(-) + 7E_u(x,y) + A_{1u}(-) + 2B_{1g}(x^2-y^2) + 2B_{2g}(xy) + 3E_g(xz,yz) + \\ + A_{1g}(x^2+y^2,z^2) + A_{2g}(-) \quad (3)$$

The symbols in parentheses denote the phonon activities in the Raman and infrared spectra. In the low-temperature chiral phase below 420 K (tetragonal space groups $P4_12_12$ and $P4_32_12$, Z = 4), the factor-group analysis of the optical phonons gives

$$\Gamma_{opt} = 9A_2(z) + 10B_1(x^2-y^2) + 8B_2(xy) + 17E(x,y,xz,yz) + 8A_1(x^2+y^2, z^2) \quad (4)$$

This means that in the chiral phase, we can expect a total of 43 Raman-active phonons, whereas in the high-temperature achiral phase, only 8 Raman-active phonons are allowed. Experimentally, we observe a much smaller number of phonons in chiral phase because many of them have low intensities or are overlapping due to finite phonon damping.

If we were measuring an oriented single crystal, we would observe only 3 modes ($2B_{1g}+ A_{1g}$) in the VV spectra and 5 other modes ($2B_{2g}+ 3E_g$) in the HV spectra at high temperatures. In our case, we observe a total of 8 modes regardless of polarization. This is caused by random orientation of crystal grains. Our spectra are practically equivalent to unpolarized spectra.

For the low temperature phase, the number of experimentally observed phonons remains practically the same as in the high-temperature phase, apart from the splitting of the soft mode. To understand this splitting, we calculated the phonon dispersion curves in both crystallographic phases with foundation (universal) machine-learning interatomic potentials (MLIPs). Several foundation machine-learning interatomic potentials were initially evaluated. Among them, PET-MAD-1.5-S [9,10,19] and SevenNet-Nano-5.5 [20,21] were selected for further analysis as both reproduced the expected qualitative difference in dynamical stability between the two phases (no imaginary phonon modes for the low-temperature phase, while predicting a phonon instability in the high-temperature phase). Trained on large and chemically diverse DFT datasets, these foundation models enable complete phonon dispersions to be calculated at a substantially lower computational cost than direct DFT calculations. However, the calculated phonon frequencies and the predicted wave vectors of the instabilities should still be interpreted with caution[22].

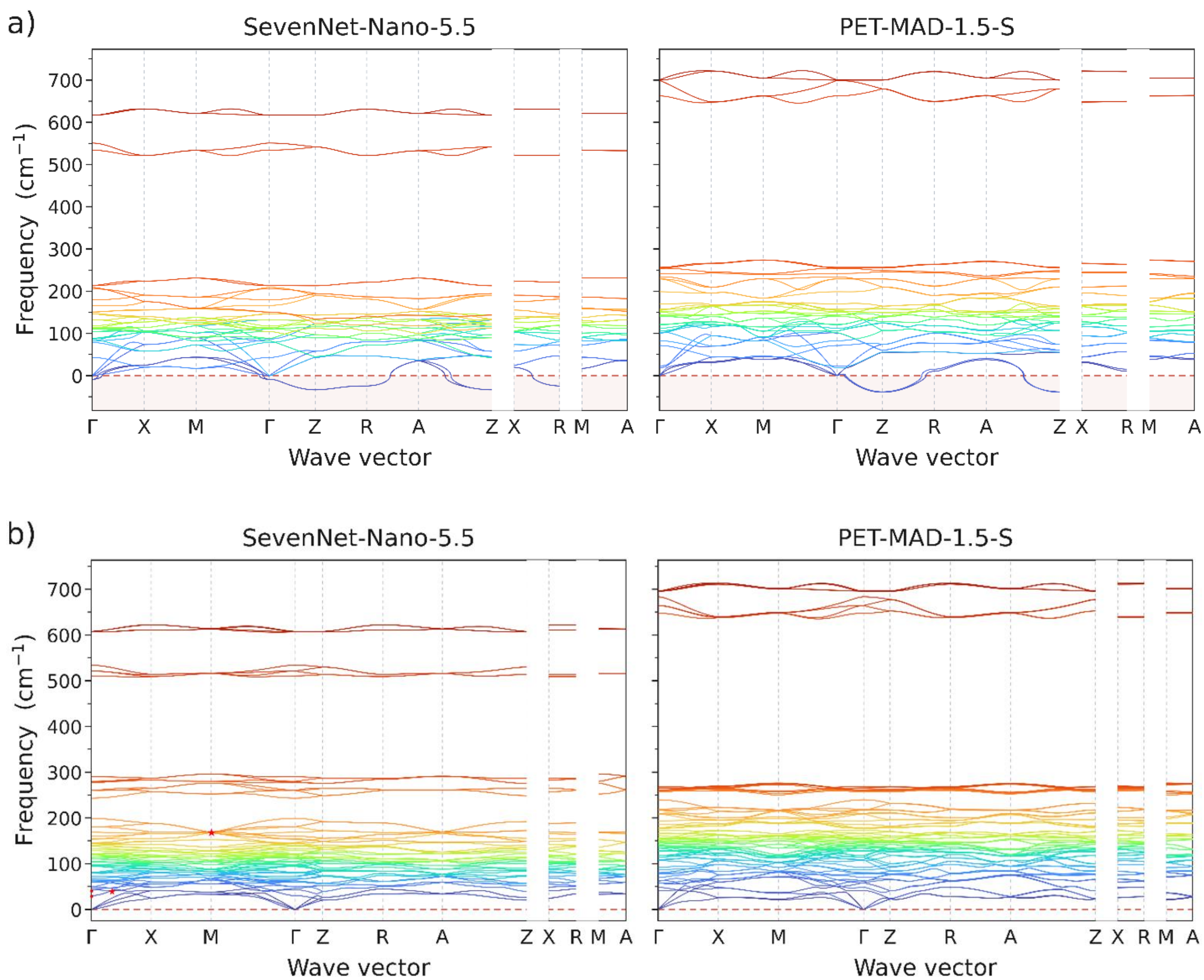

FIG. 6. Computed phonon dispersion branches in (a) high-temperature achiral phase with $P4_2/mnm$ symmetry and (b) low-temperature chiral phases with $P4_12_12$ space group. The calculations were performed with SevenNet-Nano-5.5 (left) and PET-MAD-1.5-S foundation MLIPs (right). The phonons for which atomic vibrations were calculated and saved as videos in the Supplementary information[31] are labeled with symbols.

The high-temperature phase exhibits phonons with negative frequencies at Z point in the BZ (Fig. 6a) by both foundation models. This indicates the instability of this phase and the existence of a soft mode with this wave vector from the BZ boundary. A similar result has already been published by Fava *et al.*[5]. The soft mode near Z point cannot be observed in Raman or IR spectra, because these techniques allow only measurement near Γ point of BZ.

The unit cell doubles in size below $T_c$, causing the folding of the BZ and multiplication of number of the phonon branches. Phonons from the Z point in the band structure flip over to the Γ point and can be activated in the spectra. In our case for the low-temperature phase, two Raman-active soft modes below 40 cm$^{-1}$ are activated. Fig. 6b shows two calculated low-frequency phonons at 41 and 48 cm$^{-1}$ with PET-MAD-1.5-S. Calculations using SevenNet-Nano-5.5 yielded the $A_1$ and $B_2$ soft modes at 30.3 and 43.1 cm$^{-1}$, respectively, in closer agreement with the experimental frequencies. The comparison with experiment demonstrates that foundation MLIPs provide an efficient alternative to DFT for screening phonon stability and calculating phonon dispersions, but their quantitative predictions (without any fine-tuning) remain model-dependent and should be validated against experimental data or first-principles calculations whenever possible.

The structural transition is governed by a two-dimensional $Z$ mode in the highly symmetric phase, whose condensation creates a right-handed or left-handed helix of atomic displacements in the chiral phase. Precisely this condensation is the source of the crystal's chirality.

After the transition, this $Z$ mode splits into $A_1$ and $B_1$ modes. The two resulting phonons thus represent fluctuations of the chiral order parameter around the chiral minimum. The $A_1$ mode can be understood as an amplitude oscillation of the magnitude of the chiral deformation, while the $B_1$ mode represents oscillations in an orthogonal direction within the space of the original two-dimensional order parameter. Both modes are therefore associated with chiral distortion of the structure. As can be seen from the selection rules in Eq. (4), both modes are active in parallel-polarized Raman spectra, which is consistent with the activation of these soft modes in our VV spectra. The softening of two or more phonons is known to occur near ferroelectric phase transitions, but in those cases, the ferroelectric soft modes are coupled due to their identical symmetry[23,24]. In our case, two soft modes in $K_3NiO_2$ have different $A_1$ and $B_1$ symmetries and soften toward the same $T_c$ temperature without being coupled.

The most important result of our Raman investigation is the confirmation of the theoretical prediction by Fava *et al.*[5] that the structural phase transition from the achiral to the handed chiral phase is of the displacive type, induced by geometric chiral soft phonon from the $Z$ point of the BZ.

Although this soft mode is Raman-inactive above $T_c$ due to its wavevector $\boldsymbol{k} = (0, 0, 1/2)$ (Raman and IR spectroscopy can see phonons only with $\boldsymbol{k} \approx 0$), it becomes Raman active below $T_c$, splitting in $A_1$ and $B_1$ components which gradually harden with cooling.

The question arises as to whether the observed soft modes in the low-temperature chiral structure are also chiral, i.e. circularly polarized with angular momentum,[25,26] as they are in the high-temperature achiral phase. Circularly polarized Raman spectroscopy is a suitable tool for identifying chiral phonons with small non-zero $\boldsymbol{k}$ wavevector. At the Γ point of the BZ, degenerate phonons can split for nonzero $\boldsymbol{k}$ due to chirality, and this splitting can be detected in circularly polarized Raman scattering. Such phonon splitting has been observed for example in α-HgS [27], α-quartz [28], chiral crystal of Te [29], and $Fe_2Mo_3O_8$ [30].

### 3.2. Circularly Polarized Raman spectroscopy and Chiral Phonon fingerprints

Circularly polarized (CP) Raman scattering measurement of $K_3NiO_2$ was performed at T = 80 K, i.e. at the lowest temperature available in our set-up, where the phonon damping is the lowest. The spectra, taken in backscattering geometry with four combinations of circularly polarized beams, are shown in Fig. 7. Surprisingly, two additional weak modes were revealed near 210 and 237 $cm^{-1}$, which indicates that the CP Raman is more sensitive to weak phonons than the linearly polarized Raman scattering. The fact that these excitations are phonons is also supported by our phonon calculations - see Fig. 6. A more detailed fit of the spectrum revealed as many as 13 phonons, i.e. 5 more than in Fig. 2, but still significantly fewer than allowed by the factor-group analysis in Eq. (4).

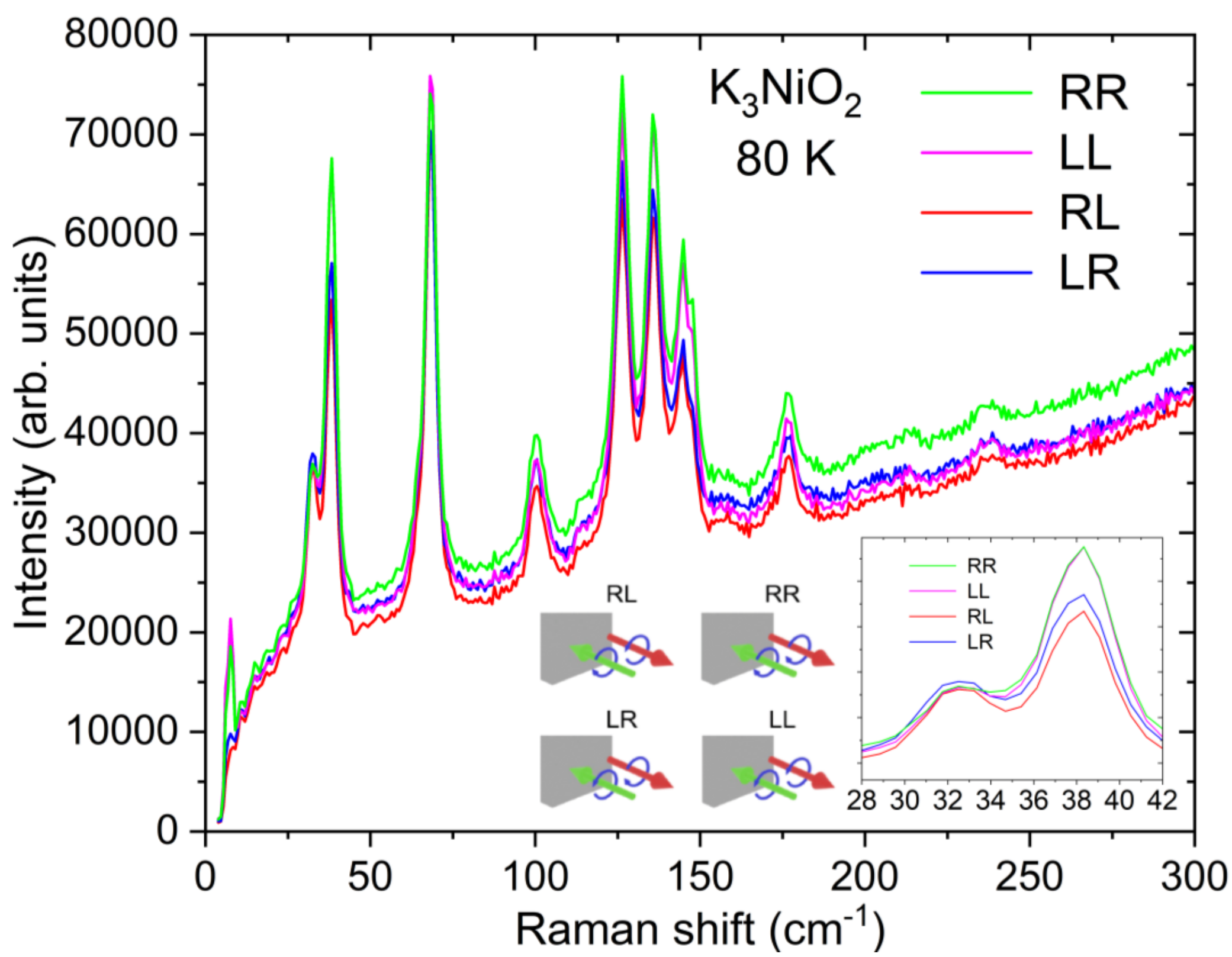

FIG. 7. Circularly polarized Raman spectra, taken at T = 80 K. Inset shows details of two lowest-frequency excitations. The following embedded image explains the notation for circularly polarized back scattering. The green and red arrows mark incident and scattered lights, respectively. The blue circular arrows indicate circular polarization, which is represented by R (right-handed) and L (left-handed).

A detailed examination of the two soft modes in the CP spectra - see the inset in Fig. 7 - revealed no additional splitting of these phonons, as had previously been observed for chiral phonons in Refs. [27,28,29,30]. This is consistent with the fact that soft modes are single-degenerate and therefore cannot split in CP Raman. However, it can be seen that the phonon bands in the parallel CP spectra (RR, LL) exhibit the same intensity, while the crossed CP spectra (RL, LR) have weaker intensity. The phonon frequency, however, remains the same for all scattering geometries. Similarly, for the other phonons observable up to 240 cm$^{-1}$, no frequency shift is seen in the various CP spectra. This could suggest that no phonons are chiral for wavevectors near Γ point of BZ. It is more likely, however, that our spectrometer's resolution (1.7 cm$^{-1}$) is not sufficient for the detection of such frequency shifts in chiral phonons. Indeed, all *E*-symmetry phonons should split near Γ point in CP Raman spectra of the chiral phase.

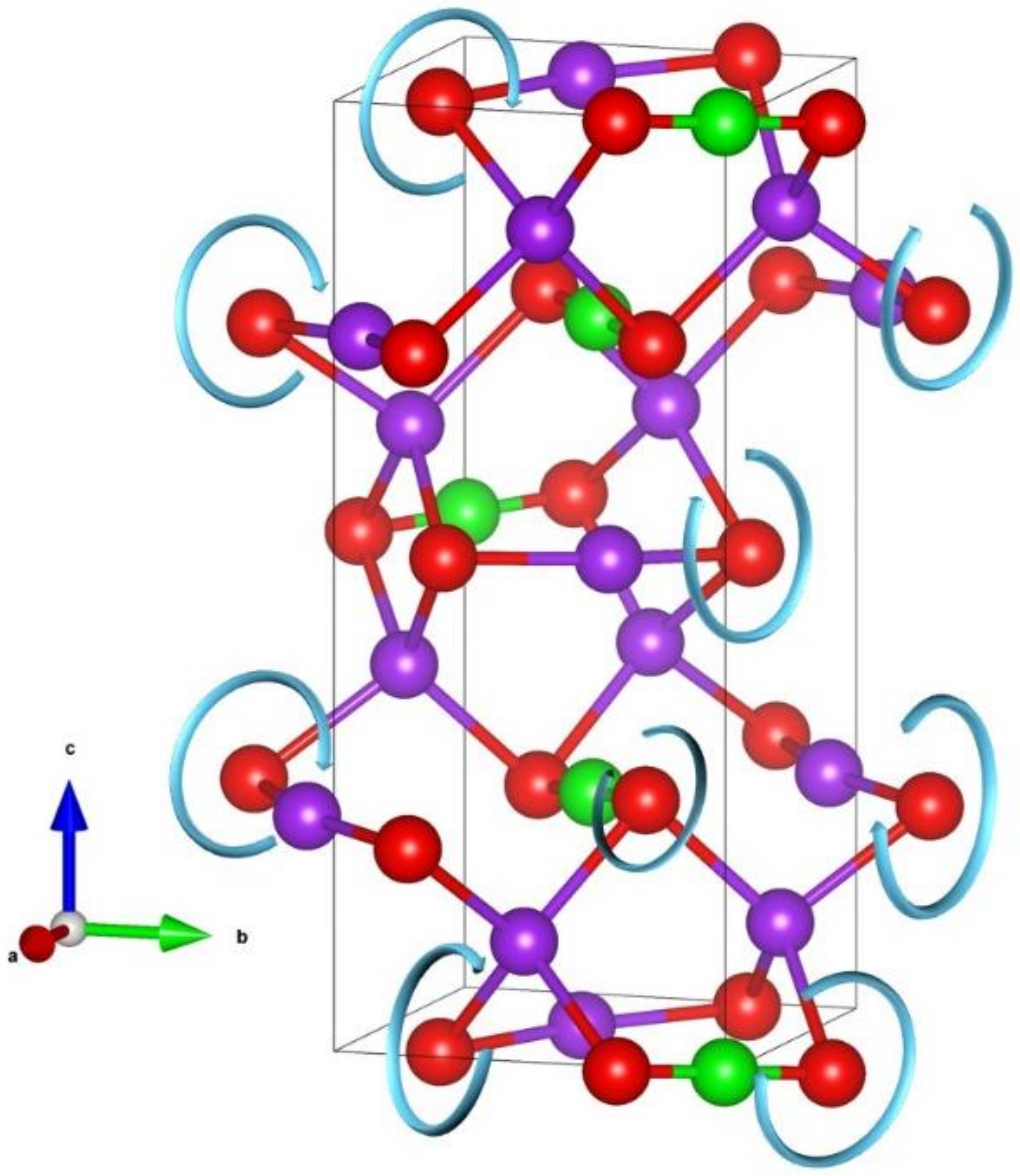

FIG. 8. A schematic representation of the low-temperature chiral phase of $K_3NiO_2$ showing circular motion of oxygen atoms, which corresponds to the Dirac phonon at the M-point of BZ with a frequency of 168 $cm^{-1}$ (see Fig. 6). A video of these vibrations is available in the SI [31]. The K, O, and Ni atoms are shown in purple, red, and green, respectively.

In the theoretical phonon dispersion curves of the low-temperature chiral phase, many degenerate phonons are visible between 50 and 300 $cm^{-1}$ (see Fig. 6), but their splitting near the Γ point is relatively weak and therefore difficult to detect in our CP Raman spectra. On the other hand, the two lowest-frequency optical phonons, calculated at 30.3 and 40.7 $cm^{-1}$, which should correspond to our observed $A_1$ and $B_1$ soft modes, show no splitting near Γ point in Fig. 6. In the SI[31], we present videos of these vibrations, demonstrating that these modes have no angular momentum. Other two videos in Supplementary Information show vibrations of topological Weyl and Dirac phonons at 43.15 and 168 $cm^{-1}$, where two and four phonon branches, respectively, intersect (see symbols in Fig. 6a). These two vibrations with $\boldsymbol{k} \neq 0$ (the latter one is from M point of BZ) exhibit a clear circular motion of the atoms and therefore have orbital angular momentum. The calculated circular motion of oxygen ions in the Dirac phonon at point M of the BZ with a frequency of 168 $cm^{-1}$ is shown schematically in Fig. 8; the full video of these motions is available in the SI [31]. The orbital momentum of the topological phonons, where the phonon branches intersect, is most pronounced, but circular motion was also observed in some other phonons for $\boldsymbol{k} \neq 0$. What is important, however, is that in the high-temperature achiral phase, no circular motion of the atoms was observed in our calculations for any wave vector $\boldsymbol{k}$. All phonons in the achiral phase are therefore achiral.

## 4. Conclusion

We synthesized $K_3NiO_2$ microcrystals with diameters of 10-20 μm using a new direct route from K, $K_2O$ and NiO. As the samples were highly unstable in air, they were measured in quartz capillaries, filled with argon. Raman spectra revealed two soft phonons that soften toward 420 K, which is the temperature of the structural transition from the chiral to the achiral structure. Above the phase transition temperature, the soft modes become Raman inactive, because they shift to the *Z* edge of the Brillouin zone. We have thus confirmed the theoretical prediction published by Fava *et al.*[5]. Our calculations of phonon dispersions throughout the Brillouin zone showed that the soft modes at the Γ point have $A_1$ and $B_1$ symmetries, so they are not degenerate; therefore, they cannot split near the Γ point and thus cannot possess angular momentum for small wavevector $\boldsymbol{k}$. We also confirmed this finding by measuring circularly polarized Raman scattering and visualizing these vibrations using theoretical calculations. However, the calculated atomic motion for certain values of $\boldsymbol{k} \neq 0$ revealed distinct circular motion of some atoms, suggesting the existence of phonon angular momentum for larger values of $\boldsymbol{k} \neq 0$ (e.g., Dirac phonon at M point in the Brillouin zone).

## Declaration of Competing Interest

The authors declare that they have no known competing financial interests or personal relationships that could have appeared to influence the work reported in this paper.

## Acknowledgements

This work was supported by the Czech Science Foundation (Project No. 24-10791S) and by the project TERAFIT – CZ.02.01.01/00/22_008/0004594 co-financed by the European Union and the Ministry of Education, Youth and Sports of the Czech Republic.

## Data availability

The data that support the findings of this article are openly available [32].

[23] J. Petzelt, G. V. Kozlov, and A. A. Volkov. Dielectric spectroscopy of paraelectric soft modes. *Ferroelectrics* **73**, 101 (1987). 10.1080/00150198708227912

[24] S. Kamba, G. Schaack, and J. Petzelt. Vibrational spectroscopy and soft-mode behavior in Rochelle salt. *Phys. Rev. B* **51**, 14998-15007 (1995).

[25] D. M. Juraschek, R. M. Geilhufe, H. Zhu, M. Basini, P. Baum, A. Baydin, S. Chaudhary, M. Fechner, B. Flebus, G. Grissonnanche, A. I. Kirilyuk, M. Lemeshko, S. F. Maehrlein, M. Mignolet, S. Murakami, Q. Niu, U. Nowak, C. P. Romao, H. Rostami, T. Satoh, N. A. Spaldin, H. Ueda, and L. Zhang. Chiral phonons. *Nature Physics* **21**, 1532 (2025). DOI: 10.1038/s41567-025-03001-9

[26] Y. Yang, Z. Xiao, Y. Mao, Z. Li, Z. Wang, T. Deng, Y. Tang, Z.-D. Song, Y. Li, H. Yuan, M. Shi, and Y. Xu. Symmetry-guided catalogue of chiral phonon materials. *Nature Physics* **22**, 884 (2026). 10.1038/s41567-026-03260-0

[27] K. Ishito, H. Mao, Y. Kousaka, Y. Togawa, S. Iwasaki, T. Zhang, S. Murakami, J.-I. Kishine, and T. Satoh. Truly chiral phonons in α-HgS. *Nature Physics* **19**, 35 (2022). DOI: 10.1038/s41567-022-01790-x

[28] E. Oishi, Y. Fujii, and A. Koreeda. Selective observation of enantiomeric chiral phonons in α-quartz. *Phys. Rev. B* **109**, 104306 (2024). DOI: 10.1103/PhysRevB.109.104306

[29] K. Ishito, H. Mao, K. Kobayashi, Y. Kousaka, Y. Togawa, H. Kusunose, J. I. Kishine, and T. Satoh. Chiral phonons: circularly polarized Raman spectroscopy and ab initio calculations in a chiral crystal tellurium. *Chirality* **35**, 338 (2023). 10.1002/chir.23544

[30] F. Wu, S. Bao, J. Zhou, Y. Wang, J. Sun, J. Wen, Y. Wan, and Q. Zhang. Fluctuation-enhanced phonon magnetic moments in a polar antiferromagnet. *Nature Physics* **19**, 1868 (2023). 10.1038/s41567-023-02210-4

[31] Supplementary information show temperature dependence of magnetic susceptibility of $K_3NiO_2$ and videa of calculated soft mode vibrations plus circular motions of atoms associated with the selected Weyl and Dirac topological phonons.

[32] Open data supporting this article is available at https://doi.org/zenodo (we will create open data at zenodo).

# *Soft and chiral phonons in chiral phase of $K_3NiO_2$*
## Supporting information

*Hong Dang Nguyen[1,2], Fedir Borodavka[1], Miroslav Lebeda[1,3], Nazar Zaremba[4], Peter Höhn[4], , Eteri Svanidze[4], Jan Drahokoupil[1,2], Aleš Vlk[1], Stanislav Kamba[1,*]*

*[1]Institute of Physics, Czech Academy of Sciences, Na Slovance 2, 182 00 Prague 8, Czech Republic*

*[2]Faculty of Nuclear Sciences and Physical Engineering, Czech Technical University in Prague, Trojanova 13, 120 00, Prague 2, Czech Republic*

*[3]Faculty of Mechanical Engineering, Czech Technical University in Prague, Technická 4, 16607 Prague 6, Czech Republic*

*[4]Max-Planck-Institute for Chemical Physics of Solids, Nöthnitzer Str. 40, 01187 Dresden, Germany*

**Magnetic properties of $K_3NiO_2$**

Temperature-dependent magnetic measurements were conducted in a Quantum Design Magnetic Properties Measurement System, equipped with a 7 T SQUID magnetometer. A collection of single crystallites was mounted inside a quartz capillary and sealed with G-Varnish glue under argon atmosphere. Magnetic susceptibility was measured at temperatures ranging from 2 to 300 K and in magnetic field of 1 T. The resulting paramagnetic behavior (i.e. increase of magnetic susceptibility $\chi$ on cooling) is consistent with previous reports[1]. The experimental $\chi(T)$ dependence (See Fig. S1) was fitted using the Curie formula $\chi(T) = {}^{C}/_{T} + \chi_{\infty}$, where the Curie constant $C$ = 0.419(4) K.emu/mol, $T$ is the absolute temperature, and $\chi_{\infty}$ = 0.01891(5) emu/mol is the background magnetic susceptibility given by small amount (on the order of ppm) of magnetic impurity.

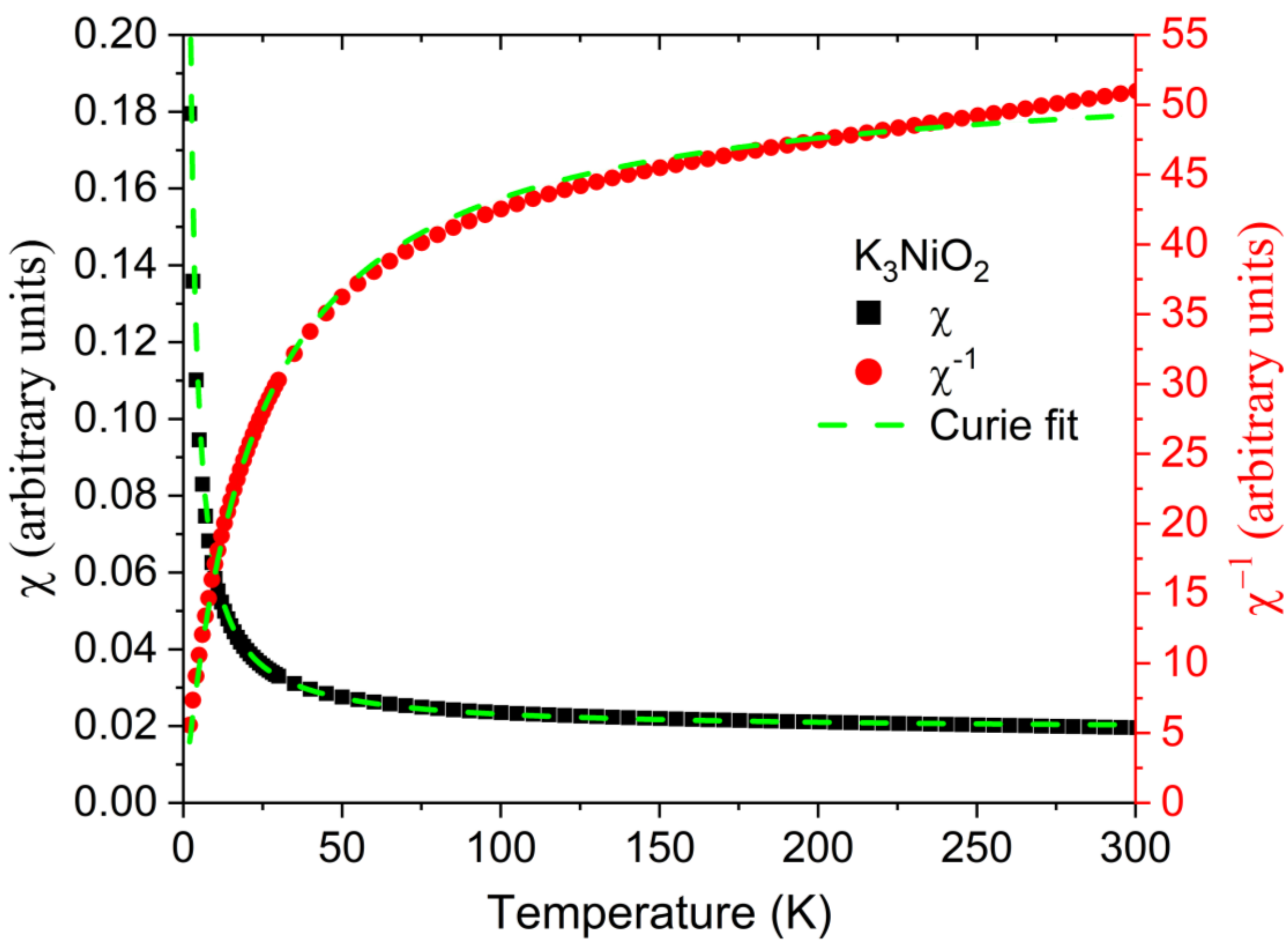

Figure S1. Magnetic susceptibility χ and its reciprocal value $\chi^{-1}$ of $K_3NiO_2$ single crystalline powder (gray/red dots) with the Curie fitting (green line) in the range of 2 K to 300 K. The sample was kept in a capillary during the measurement and showed paramagnetic behavior. Due to the difference between the impurities in our sample and in the sample from Ref. [1], the values may be slightly different. However, the overall paramagnetic behavior is still well-pronounced in both studies.

**Lattice vibrations visualizations**

We calculated the lattice vibrations of $K_3NiO_2$ and visualized them using the TSS Physics software – Visualization of phonons (https://henriquemiranda.github.io/phononwebsite/phonon.html). We have selected and posted four videos of crystal lattice vibrations for our readers. Frequencies and ***k*** vectors of these modes are marked in Fig. S2. Two of the videos show the vibrations of atoms at 30.3 and 40.7 $cm^{-1}$ (red dots) at the Γ point of BZ. These are the soft phonons active in Raman scattering. These videos clearly demonstrate that these soft phonons exhibit no orbital momentum. Other two videos show vibrations of topological Weyl and Dirac phonons at 43.15 and 168 $cm^{-1}$ (red stars), where two and four phonon branches, respectively, intersect. These two vibrations with $\boldsymbol{k} \neq 0$ (the latter one is from M point of BZ) exhibit a clear circular motion of the atoms and therefore have orbital angular momentum.

The orbital momentum of these phonons is most pronounced, but circular motion was also observed in some other phonons for $\boldsymbol{k} \neq 0$. What is important, however, is that in the high-temperature achiral phase, no circular motion of the atoms was observed for any wave vector k. All phonons in the achiral phase are therefore achiral.

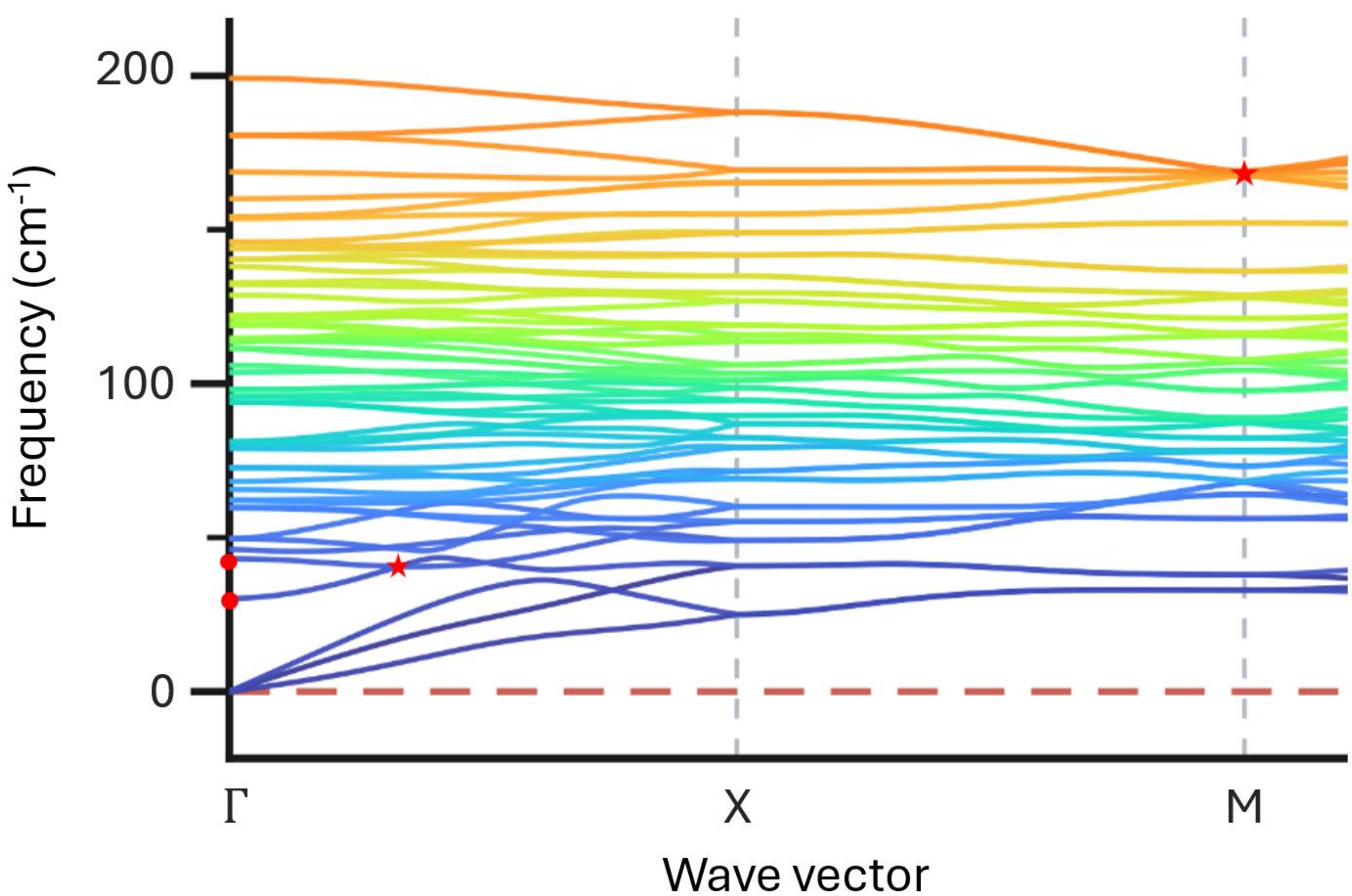

Figure S2. A selected portion of the phonon dispersion branches in the BZ of chiral phase. The phonons for which videos with vibrations have been uploaded are marked. The red dots represent two soft modes at 30.3 $cm^{-1}$ and 43.15 $cm^{-1}$ at the Γ point. Two red stars denote the topological Weyl phonon at 40.71 $cm^{-1}$, and the other one at 168 $cm^{-1}$ represents the topological Dirac phonon.

**List of video files:**

30.3cm-1_Gamma.gif
40.71cm-1_Weyl-topological_phonon.gif
43.15cm-1_Gamma.gif
168cm-1_Dirac-topological_phonon.gif